\documentclass{aastex61}

 \usepackage{CJK}
\begin{document}

 \begin{CJK*}{UTF8}{gbsn}

\title{Simultaneous existence of the ocsillations, counterstreaming flows and mass injections in solar quiescent prominences}

\correspondingauthor{Xiaoli Yan}
\email{yanxl@ynao.ac.cn}

\author{Xiaoli Yan}
\author{Zhike Xue}
\author{Jincheng Wang}
\affiliation{Yunnan Observatories, Chinese Academy of Sciences, Kunming 650216, People's Republic of China.}
\affiliation{Yunnan Key Laboratory of Solar Physics and Space Science, Kunming, 650216, People's Republic of China.}

\author{P.F. Chen}
\affiliation{School of Astronomy and Space Science, Nanjing University, Nanjing, 210023, People's Republic of China.}
\affiliation{Key Laboratory of Modern Astronomy and Astrophysics, Ministry of Education, Nanjing, 210023, People's Republic of China.}

\author{Kaifan Ji}
\affiliation{Yunnan Observatories, Chinese Academy of Sciences, Kunming 650216, People's Republic of China.}

\author{Chun Xia}
\affiliation{School of Physics and Astronomy, Yunnan University, Kunming, 650050, Peopleʼs Republic of China.}

\author{Liheng Yang}
\author{Defang Kong}
\author{Zhe Xu}
\author{Yian Zhou}
\affiliation{Yunnan Observatories, Chinese Academy of Sciences, Kunming 650216, People's Republic of China.}
\affiliation{Yunnan Key Laboratory of Solar Physics and Space Science, Kunming, 650216, People's Republic of China.}

\author{Qiaoling Li}
\affiliation{School of Physics and Astronomy, Yunnan University, Kunming, 650050, Peopleʼs Republic of China.}

\begin{abstract}
Solar prominences are very spectacular structures embedded in the tenuous and hot solar corona. The counterstreaming flows, a common feature in solar quiescent prominences, have been discovered for more than twenty years. However, the mechanism driving the counterstreaming flows is still elusive. To unveil the nature of this phenomenon, we analyzed the data of a quiescent prominence observed by the New Vacuum Solar Telescope (NVST), the Interface Region Imaging Spectrograph (IRIS), and the Solar Dynamical Observatory (SDO). It is found that there is a distinct longitudinal oscillation of prominence plasma along the higher part of the prominence spine in H$\alpha$ observations. The oscillation period is approximately 83 minutes and the amplitude is about 32 Mm. The counterstreaming flows are dominant in the middle part of the prominence spine. The velocities of the counterstreaming flows range from about 4 km s$^{-1}$ to 11 km s$^{-1}$. Moreover, the intermittent mass flows with the upward plumes from the top of the bubbles and tornado-like barbs are observed to be injected into the lower part of the prominence spine from the lower atmosphere. The velocities of these injected mass flows range from about 3 km s$^{-1}$ to 30 km s$^{-1}$. Some injected mass flows exhibit redshifted Doppler signals, while others exhibit blueshifted signals. Based on these high resolution observations, it is found that different parts of the prominence spine exhibit the different dynamic characteristics. These results further advance the understanding of the ubiquitous counterstreaming flows in solar quiescent prominences.

\end{abstract}
 

\section{Introduction}\label{sec:introduction}
Solar quiescent prominences are magnetized structures containing relatively cool (T $<$ 2 $\times$ 10$^4$ K) and dense materials (10$^9$ $<$ n$_e$ $<$ 10$^{11}$ cm$^{-3}$) suspended in the hot corona (Priest 1989; Tandberg-Hanssen 1995). In comparison to active-region prominences, quiescent prominences are long-lived structures, enduring for several days or even months (Parenti 2014; Vial \& Engvold 2015). When appearing at the solar limb, they typically consist of three linked substructures: a long spine, bubbles, and legs. The spine, the highest and relatively horizontal axial part of prominences, is composed of many thin threads at a certain angle relative to the axis of prominences, as revealed by high-resolution observations (Lin et al. 2005). The legs, also known as barbs when observed on the solar disk, are short protrusions from the spine, similarly consisting of many thin threads. Bubbles, distinct from the other substructures, resemble cavities with a semi-circular shape under the spine.

Early spectroscopic observation indicated that solar prominences, which are nearly quasi-static as a whole, are actually dynamic, with the coexistence of redshifted and blueshifted motions in active region filament threads (Schmieder et al. 1991). These motions were later identified in the quiescent prominence by using H$\alpha$ off-band data from Big Bear Solar Observatory, termed as the counterstreaming flows (Zirker et al. 1998), existing in adjacent thin threads of the spine and barbs. Increasing observational evidence confirmed the common occurrence of counterstreaming flows in quiescent prominences (Schmieder et al. 2008; Shen et al. 2015; Diercke et al. 2018). The velocities of the counterstreaming flows range from about 8 km s$^{-1}$ to 15 km s$^{-1}$ (Lin et al. 2003; 2005).  Similar to the quiescent prominences, active-region filaments also possess counterstreaming flows (Deng et al. 2002; Alexander et al. 2013; Wang et al. 2018). Except for the coexistence of the counterstreaming flows in prominences, periodic back and forth mass movements in the prominences/filament were also reported by Jing et al. (2006). Furthermore, longitudinal and transverse oscillations are important characteristics in the dynamic features of solar prominences (Chen et al. 2008, 2020; Ning et al. 2009a, 2009b; Zhang et al. 2012; Zhou et al. 2021; Okamoto et al. 2007, 2015). The triggering mechanism of some prominence oscillations are found to be associated with the interaction between the prominences and wave or jet activities (Luna et al 2014; Zhang et al. 2017, 2020, 2024).


While ground-based and space-borne solar telescopes have revealed more fine structures of solar prominences (Yan et al. 2015; Li et al. 2018a, 2018b; Awasthi et al.2019; Bi et al. 2020; Xue et al. 2021; Wang et al. 2022; Song et al. 2024), the driving force behind counterstreaming flows in quiescent prominences remains unclear (Chen et al. 2020). Simulation results suggest that turbulent heating at the solar surface, leading to the random evaporation of materials from the solar surface to the corona, can generate the formation of sparse threads in the solar corona (Zhou et al. 2020). Once formed, the turbulent heating would also drive ceaseless oscillations of the prominence threads, not in phase with each other, leading to counterstreaming flows in prominences or filaments (Chen et al. 2014; Zhou et al. 2020). Additionally, network jets are identified as drivers of counterstreaming flows in a small quiescent filament on the solar disk (Panesar et al. 2020).  

To uncover the nature of counterstreaming flows in solar prominences, we utilize the high resolution multi-wavelength imaging and spectral data from the New Vacuum Solar Telescope (NVST), the Interface Region Imaging Spectrograph (IRIS), and the Solar Dynamical Observatory (SDO) to study the evolution of a solar prominence on 2016 August 17. The related content is organized as follows: The observations and methods are described in Section 2. The results are presented in Section 3. The conclusions and discussions are given in Section 4.

\section{Observations and methods}\label{sec:observations}
The NVST observed a prominence located above the west limb of the Sun On 2016 August 17. We used the H$\alpha$ line-center, and the H$\alpha$ off-band images acquired at 6562.8~{\AA}, 6562.8 $\pm$ 0.4~{\AA} , and 6562.8 $\pm$ 0.7~{\AA} observed by the NVST to study the oscillation, mass injection, and plume evolution of the prominence on 2016 August 17. The spatial resolution of these images is 0.$^\prime$$^\prime$3. The time cadence is $\sim$76 s (Liu et al. 2014; Yan et al. 2020). These data are first calibrated from Level 0 to Level 1 with the dark current subtracted and flat-field corrected. Then, the speckle masking method is used to calibrate the images from Level 1 to Level 1+ (Xiang et al. 2016). We coaligned the NVST H$\alpha$ line-center, and off-band images by using the method of Ji et al. (2019) and Cai et al. (2024). The co-alignment accuracy is about 0.$^\prime$$^\prime$1.

Three wavelength EUV images (304~\AA, 171~\AA, and 211~\AA\ channels) observed by the Atmospheric Imaging Assembly (AIA) (Lemen et al. 2012) on board the Solar Dynamic Observatory (SDO) (Pesnell et al. 2012) are used to illustrate the prominence. These images have a spatial resolution of $1.^{\prime\prime}5$ and a cadence of 12 s. 

The IRIS performed the observations at Mg II h 2803.5 from 07:49:18 UT to 09:14:56 UT with 5 large coarse raster scans, and each scan had 64 $\times$ 1$^\prime$$^\prime$ raster steps (De Pontieu et al. 2014). The pixel size along the raster slit is 0.33$^\prime$$^\prime$. The field of view is 63$^\prime$$^\prime$ $\times$ 119$^\prime$$^\prime$.

The Dopplergrams are constructed by using the following equation (Langangen et al. 2008), D = (B - R)/(B + R)
where B and R represent the H$\alpha$ blue-wing (-0.4 and -0.7 \AA) and red-wing (+0.4 and +0.7 \AA) images, respectively. Note that the Dopplergrams just show the Doppler shift signals.

\section{Results}\label{sec:results}
\subsection{Appearance of the prominence}
On 2016 August 17, a prominence appeared above the west limb of the Sun. Figure 1 illustrates the appearance of the filament in EUV images (304 \AA, 171 \AA, 211 \AA) at 06:17:58 UT (Figures 1a--1c) and H$\alpha$ center images from 06:14:08 UT to 07:43:09 UT (Figures 1d--1f). For a clearer presentation of the prominence structures, the original images are rotated 59 degrees counterclockwise. The prominence displays emissive structures in 304 \AA\ image (Figure 1a) and H$\alpha$ center images (Figures 1d--1f) and absorptive structures in the 211 \AA\ image (Figure 1c). The coexistence of emissive and absorptive structures can be seen in the 171 \AA\ image (Figure 1b). The spine of the prominence is nearly parallel to the solar limb. The Supplementary Movie 1 provides an animation of the 304 \AA, 171 \AA, and 211 \AA\ observations, showcasing the evolution of the prominence. Seen from H$\alpha$ center images, this prominence comprises of a spine (pink arrow), several bubbles (blue arrows), and legs (red arrows)  shown in Figure 1d. As the usual appearance of solar quiescent or polar crown prominences, this prominence appear to be feet (barbs) dominated and their barbs appear as dense pillars, and the main body/spine of the filament is weakly visible (e.g. Schmieder et al. 2010, Dud{\'\i}k et al. 2012, Panesar et al. 2014). The appearance of the present prominence is similar to the prominence shown in Dud{\'\i}k et al. (2012) and Shen et al. (2015). The blue box in Figure 1a outlines the field of view of NVST. Despite the NVST's limited field of view, approximately 180$^\prime$$^\prime$ $\times$ 180$^\prime$$^\prime$, it successfully captures most of the prominence. 

\subsection{Counterstreaming flows along the prominence spine}
Counterstreaming flows are commonly observed in solar prominences. To analyze the counterstreaming flows along the spine of the prominence, we created time-distance diagrams from 06:00:09 UT to 09:00:43 UT along slit A0-B0, as marked in Figure 1e. Additionally, we generated 11 time-distance diagrams that are parallel to the slit A0-B0 along the prominence spine and found that the counterstreaming flows along the spine were most prominent in the central region of the prominence.

Using H$\alpha$ center images, H$\alpha$ off-band images, and reconstructed Dopplergrams, we confirmed the presence of counterstreaming flows along the prominence spine. Figure 2a illustrates the counterstreaming flows observed in the H$\alpha$ center images. The dashed pink and cyan lines indicate flows in opposite directions, with projection velocities ranging from approximately 4 km s$^{-1}$ to 11 km s$^{-1}$. Figure 2b presents a time-distance diagram derived from Dopplergrams at H$\alpha$ +0.4 \AA, where the solid pink and cyan lines similarly highlight counterstreaming flows with projection velocities between 4 km s$^{-1}$ and 10 km s$^{-1}$, which is similar to the results obtained by Schmieder et al. (2010).

We also used H$\alpha$ off-band -0.4 \AA\ images to generate time-distance diagrams along slit A0-B0. These diagrams revealed unidirectional flows moving from A0 to B0, as shown in Figure 2c. Conversely, using H$\alpha$ off-band +0.4 \AA\ images, the diagrams displayed unidirectional flows moving from B0 to A0, as shown in Figure 2d. The projection velocities of these unidirectional flows ranged from approximately 4 km s$^{-1}$ to 10 km s$^{-1}$.

\subsection{Prominence oscillation}
The primary prominence spine, observed comprehensively from approximately 06:00 UT to 09:00 UT on 2016 August 17, is selected for a detailed analysis of oscillations and mass flows. Figures 3a1--3c1 display the H$\alpha$ center images, Dopplergrams reconstructed using $\pm$ 0.4~{\AA} and $\pm$ 0.7~{\AA} images at 06:42:06 UT, respectively. The field of view of Figures 3a1--3c1 is outlined by the red box in Figure 1e. Blueshifted and redshifted Doppler signals are represented by blue and red colors, respectively. To analyze the oscillation along and across the spine of the prominence, two straight slices perpendicular to each other are chosen to create the time-distance diagrams. Slices A1--B1, A2--B2, and A3--B3 run along the spine, while slices C1--D1, C2--D2, and C3--D3 are perpendicular to the spine (see the white lines in Figure 3a1 and the black lines in Figures 3b1 and 3c1). This selection allows us to derive longitudinal and transverse oscillations of the prominence, respectively. The time-distance diagrams along the slices A1--B1, A2--B2, and A3--B3 are shown in Figures 3a2, 3b2, and 3c2, respectively. In the time-distance diagram using the H$\alpha$ center images (see Figure 3a2), the direction of the mass flows changes three times between 10 Mm to 35 Mm along the slice. The velocities of the oscillation range from about 7 km s$^{-1}$ to 9 km s$^{-1}$. The associated Supplementary Movies 2, 3, and 4 provide animations of H$\alpha$ center and the dopplergrams at H$\alpha$ $\pm$ 0.4~{\AA} and $\pm$ 0.7~{\AA}.

Examining the time-distance diagrams along slices A2--B2 and A3--B3 using H$\alpha$ Dopplergrams, the alternating blue and red shifts become evident along the prominence spine (see Figures 3b2 and 3c2) in the dopplergrams at H$\alpha$ $\pm$ 0.4~{\AA}, and $\pm$ 0.7~{\AA} as well as the intensity time-distance diagram shown in Figure 3a2. The vertical green lines artificially separate the red shift from the blue shift in Figures 3b2 and 3c2. The interval time of the redshifted signal and blueshifted signal is approximately 40 minutes (38.1 and 43.2 minutes for the two off bands, respectively). These observations confirm the presence of longitudinal oscillations in the prominence spine, with an oscillation amplitude of about 32 Mm along slices A1--B1, A2--B2, and A3--B3. Furthermore, the velocities of the blue shift and the red shift along the spine of the prominence range from about 11 km s$^{-1}$ to 17 km s$^{-1}$, similar to the velocities of the flows along the filament thread structures on the solar disk observed by Lin et al. (2005) and derived from simulation (Zhou et al. 2020). It is noteworthy that the oscillation signals are more pronounced in the dopplergrams than in the H$\alpha$ filtergrams.

In addition to the oscillations along the spine of the prominence, numerous small-amplitude transverse oscillations of the prominence threads, perpendicular to the spine, are observed. These oscillations along the slice C1--D1 are outlined by the curved white lines in Figure 3a3. The curved lines in Figure 3a3 delineate the small amplitude oscillation of the prominence plasma. The oscillation periods range from about 12 minutes to 17 minutes, with an amplitude of approximately 3 Mm. The time-distance diagrams along the slices C2-D2 and C3-D3, using the dopplergrams at H$\alpha$ $\pm$ 0.4~{\AA}, and $\pm$ 0.7~{\AA}, are presented in Figures 3b3 and 3c3, respectively. The alternating appearance of blueshifted and redshifted signals at the higher part of the prominence spine is evident in the time-distance diagrams, denoted by the pink and black arrows. The blue and the red shifts indicate the existence of the mass oscillation along the higher part of the prominence spine as observed in Figs. 3b3 and 3c3. At the lower spine of the prominence, both the blue and the red shifts coexist, with the Doppler shifts exhibiting randomness. This may be attributed to the Doppler signals in the lower part of the prominence spine being influenced by injected mass flows from the lower atmosphere, which is to be discussed in the following section.

To precisely determine the period of the longitudinal oscillations, intensity profiles along the yellow lines in Figures 3a2, 3b2, and 3c2 are plotted in Figures 4a, 4b, and 4c. Wavelet analysis is employed to extract the periods of the longitudinal oscillation. The oscillation period, derived from the normalized intensity of the H$\alpha$ center image, is 80.3 minutes, while from the dopplergrams at $\pm$ 0.4~{\AA} and $\pm$ 0.7~{\AA}, it is 83.1 minutes (see the second column of Figure 3). This duration is slightly longer than the period of 40--60 minutes derived from the tornado-like prominence (Schmieder et al. 2017). 

IRIS also observed a portion of this prominence spine. The intensity images obtained at Mg II h 2803.5 \AA\ are presented in Figures 5a and 5d. The field of view of Figures 5a and 5d is outlined by the white box in Figure 1b. The corresponding Doppler velocities derived from this spectral line are shown in Figures 5b and 5e. It is evident that the Doppler signals in the black circles change from blueshift to redshift between 07:49 UT to 08:23 UT. Profiles of Mg II h line at the asterisks in Figures 5a and 5d are shown in Figures 5c and 5f, with the asterisks located within the black circles. The Gaussian fitting method was employed to derive the velocities of the prominence indicated by the asterisks. The Doppler velocities derived from the spectra are -8.26 km s$^{-1}$ at 07:54:57 UT and 5.02 km s$^{-1}$ at 08:29:18 UT, respectively. Considering the horizontal velocities along the prominence spine, the total velocities of mass flows in the prominence range from 12 km s$^{-1}$ to 17.7 km s$^{-1}$. Due to the lower temporal resolution of the IRIS spectra, only five dopplergrams are obtained from 07:49:18 UT to 09:14:56 UT. Nevertheless, the IRIS observations clearly confirm the NVST observations, indicating the presence of a large-scale longitudinal oscillation in the higher part of the prominence spine. 


\subsection{Mass injection}
Through analyzing the evolution of the prominence in H$\alpha$ observations, two types of mass injections are identified. The first type is associated with upward plumes (Berger et al. 2008, 2010, 2011), while the second is linked to tornado-like barbs (Li et al 2012, Su et al 2012, Panesar et al 2013, Panasenco et al 2014).

To illustrate the first type of mass injection, we selected a section of the prominence marked by the red box in Figure 1f. Figures. 6a1, 6b1, and 6c1 to depict the H$\alpha$ center image, and the dopplergrams at $\pm$ 0.4~{\AA}, and $\pm$ 0.7~{\AA}, respectively. Before we see the mass injection, this part of the prominence consists of a prominence spine, a small bubble, and a larger bubble outlined by the dotted red line in Figure 6a1. The mass injection originates from the top of the larger bubble, specifically at the location marked as the letter E1 in Figure 6a1. Several uninterrupted plumes rise into the prominence spine, and plasma is injected into the prominence along the boundary of the larger bubble from the lower atmosphere, accompanying the rising plumes from 06:00:09 UT to 07:03:43 UT. Once injected nearly vertically, the direction of mass flows becomes horizontal, and most of the injected mass descends along one leg of the arched bubble. The remaining mass participates in the longitudinal oscillation of the prominence. The evolution of the prominence is visualized in the associated Supplementary Movie 2. Unlike the dark upflows (also known as plumes) studied by Berger et al. (2008, 2010, 2011), which originate from the top the prominence bubbles, the rising of the plumes in this prominence carries mass from the lower atmosphere along the boundary of the bubbles into the prominence.

The Doppler diagrams, created using the H$\alpha$ $\pm$ 0.4~{\AA} and $\pm$ 0.7~{\AA} images, reveal red-shifted signals (see Figures 6b1 and 6c1). To determine the upward velocity of the mass injection, we use slices E1--F1, E2--F2, and E3--F3 to generate time-distance diagrams. From Figure 6a2, the velocities of the mass injection in the H$\alpha$ center observation range from about 22 km s$^{-1}$ to 30 km s$^{-1}$ along the slices E1--F1. The velocities obtained in this prominence are comparable to the velocities of the upflows obtained by Berger et al. (2008, 2010) from the Hinode data. It is noteworthy that the velocities align with the direction of the upward plumes. In addition to the H$\alpha$ center observations, we also obtained the H$\alpha$ off-band observations. It is found that the mass injection consistently exhibits redshifted signals in $\pm$ 0.4~{\AA} and $\pm$ 0.7~{\AA} Dopplergrams. The velocities of the mass injection, calculated via the dopplergrams at $\pm$ 0.4~{\AA}, and $\pm$ 0.7~{\AA} along the slices E2--F2 and E3--F3, are approximately 16 km s$^{-1}$ (see Figures 5b2 and 5c2). 

To showcase the second type of mass injection, we select another region of the prominence marked by the cyan box in Figure 1f. The corresponding H$\alpha$ center image, dopplergrams at $\pm$ 0.4~{\AA}, and $\pm$ 0.7~{\AA} are presented in Figures 6a3, 6b3, and 6c3, respectively. A tornado-like barb, indicated by the red arrow, is located at the center of Figure 6a3. From 07:03:43 UT to 09:01:59 UT, continuous mass injection is observed from the footpoint of the tornado to the spine of the prominence. Along the silces G1--H1, G2--H2, and G3--H3, three time-distance diagrams are created using the H$\alpha$ center images, dopplergrams at $\pm$ 0.4~{\AA}, and $\pm$ 0.7~{\AA}. The mass injection velocity is then calculated. The time-distance diagram along slice G1--H1 is shown in Figure 6a4, where the velocities of the mass injection range from around 6 km s$^{-1}$ to 9 km s$^{-1}$.  Along the slices G2--H2 and G3--H3, the velocities range from about 11 km s$^{-1}$ to 15 km s$^{-1}$ (see Figs. 6b4 and 6c4). The dopplergrams exhibit the red-shifted signals during the mass injection.

The first type of mass injection also occurred in the left part of the prominence (see the cyan box in Figure 1e). The process of mass injection is very similar to that presented in Figures 7a1, 7b1, and 7c1. This part of the prominence is also composed of a part of the prominence spine, a small bubble, and a larger bubble outlined by the dotted red line in Figure 7a1. The evolution closely resembles that of the mass injection shown in Figures 6a1-6c1. The mass along the boundary of the larger bubble is injected into the prominence from the lower atmosphere, accompanying the rising of the plumes from 07:02:27 UT to 09:01:59 UT. Moreover, the plumes destroyed the structure of the large bubble and then the mass is injected into the prominence from the top of the larger bubble (the middle of the slit I1-J1). Three slices I1--J1, I2--J2, and I3--J3 in the H$\alpha$ center image, Dopplergrams at $\pm$ 0.4~{\AA}, and $\pm$ 0.7~{\AA}, are used to calculate the velocities of the mass injection. The velocities range from about 5 km s$^{-1}$ to 11 km s$^{-1}$. In contrast to red-shifted upflows shown in Figure 6, the Dopplergrams exhibit the blue-shifted signals during the mass injection.

\section{Conclusions and discussions}

Based on NVST and IRIS observations, significant longitudinal mass oscillations were detected along the upper section of the prominence spine in both H$\alpha$ line center and off-band observations. The oscillation period is approximately 83 minutes, with an amplitude of around 32 Mm. The counterstreaming flows are prominent in the middle part of the prominence along the spine. The projection velocities of the counterstreaming flows range from about 4 km $s^{-1}$ to 11 km $s^{-1}$. Additionally, smaller amplitude transverse oscillations (about 2 Mm) with shorter periods (12-17 minutes) were observed in the prominence threads. Intermittent mass flows were noted, initially moving along the prominence bubble from their bases and then being injected into the upper part of the prominence. Some mass was also seen being injected into the lower section of the prominence spine via tornado-like barbs. The velocities of these injected mass flows ranged from approximately 3 km $s^{-1}$ to 30 km $s^{-1}$, displaying different Doppler signals. Some of the injected mass flows showed redshifted Doppler signals, while others exhibited blueshifted signals. 

The phenomenon of counterstreaming flows in solar quiescent prominences has puzzled solar physicists for more than twenty years. Through the analysis of high resolution observational data from NVST, the large-scale mass oscillation along the higher spine of the prominence is found by using H$\alpha$ line center images, and Dopplergrams at H$\alpha$ $\pm$ 0.4~{\AA}, and $\pm$ 0.7~{\AA}. The longitudinal oscillation in the higher part of the prominence spine is more pronounced than that in the lower part of the prominence spine. The obtained oscillation period and the velocity in the higher part of the prominence closely match the simulation results (Zhou et al. 2020). However, the longitudinal oscillation of the lower part of the prominence spine appears to be random, likely influenced by mass injections from the lower atmosphere. Chen et al. (2014) suggested that counterstreaming flows result from the combination of filament thread longitudinal oscillations and alternating unidirectional flows, a proposition later supported by simulations (Zhou et al. 2020). In our observations, the longitudinal oscillation of the prominence plasma is significant in the higher part of the prominence spine, while the lower part exhibits both longitudinal oscillation and mass motions accompanying upward plumes and tornado-like barbs.  

Our observational results are consistent with the two popular formation mechanisms of solar prominences, i.e., the chromospheric evaporation--coronal condensation model and the direct injection model. It seems that even after a prominence is formed, the two processes are still working, which disturb the prominence to oscillate longitudinally, forming counterstreaming flows. While Zhou et al. (2020) focused on the counterstreaming flows due to random chromospheric evaporation, we found that continuous mass injection would also contribute to the counterstreaming flows. The observations reveal that when injected mass encounters oscillating prominence mass, the Doppler signals become complex, explaining the irregularities in the lower part of the prominence. Consequently, counterstreaming flows, particularly in the lower part of the prominence, arise from the interplay between prominence longitudinal oscillations and mass injections. When observing Dopplergrams from the top of the prominence, the coexistence of blueshifted and redshifted signals along the prominence spine results from the superimposition of injected mass flows and mass longitudinal oscillations—this is the inherent nature of counterstreaming flows.

It is noted in passing that two types of mass injection are observed in the prominence: injections with upward plumes and tornado-like barbs. Some injected mass flows exhibit blueshifted Doppler signals, while others exhibit redshifted signals. This is probably due to that the injection sites are at opposite sides of the prominence, i.e., those closer to the observers generate redshifts and those behind the prominence generate blueshifts.

Acknowledgments: We sincerely appreciate the suggestions and comments put forward by the reviewer that helped to improve this paper. We would like to thank the NVST, IRIS, and SDO teams for high-cadence data support. This work is sponsored by the National Science Foundation of China( NSFC) under the numbers 12325303, 12473059, 12373115, 12203097, 11973084, and 12127901, the Strategic Priority Research Program of the Chinese Academy of Sciences, Grant No. XDB0560000, Yunnan Key Laboratory of Solar Physics and Space Science under the number 202205AG070009, Yunnan Fundamental Research Projects under the numbers 202301AT070347, 202301AT070349. 


 \begin{figure*}
  \centering
   \includegraphics[width=16cm]{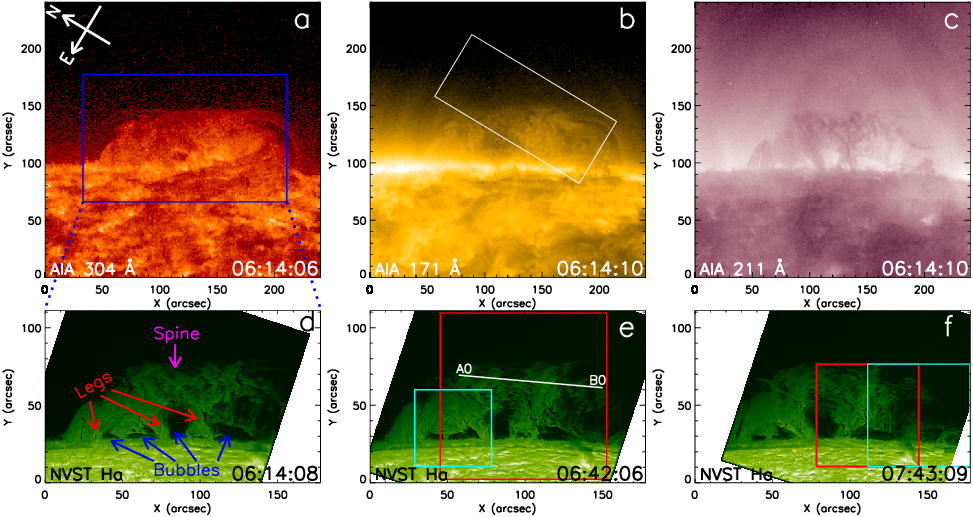}
\caption{{\bf Appearance of a quiescent prominence.} Panels (a)-(c): The quiescent prominence acquired at 304 \AA, 171 \AA, and 211 \AA\ observed by SDO/AIA at 06:14 UT on 2016 August 17. Note that the original images are rotated 59 degrees counterclockwise. Panels (d)-(f): The H$\alpha$ line center images observed by the NVST to show the evolution of the quiescent prominence from 06:14 UT to 07:43 UT. The blue box in Figure 1a outlines the field of view of the NVST. The white box in Figure 1b outlines the field of view of the IRIS. The cyan and red boxes in panels (e) and (f) outline the scope of Figs. 2, 5, and 6. The red, blue, and pink arrows in Figure 1d denote the legs, the bubbles, and the spine of the prominence, respectively. The slit A0-B0 in Figure 1e denotes the position of the time-distance diagram shown in Figure 2. 
}
     \label{fig1}
   \end{figure*}
   
      \begin{figure*}
  \centering
   \includegraphics[width=16cm]{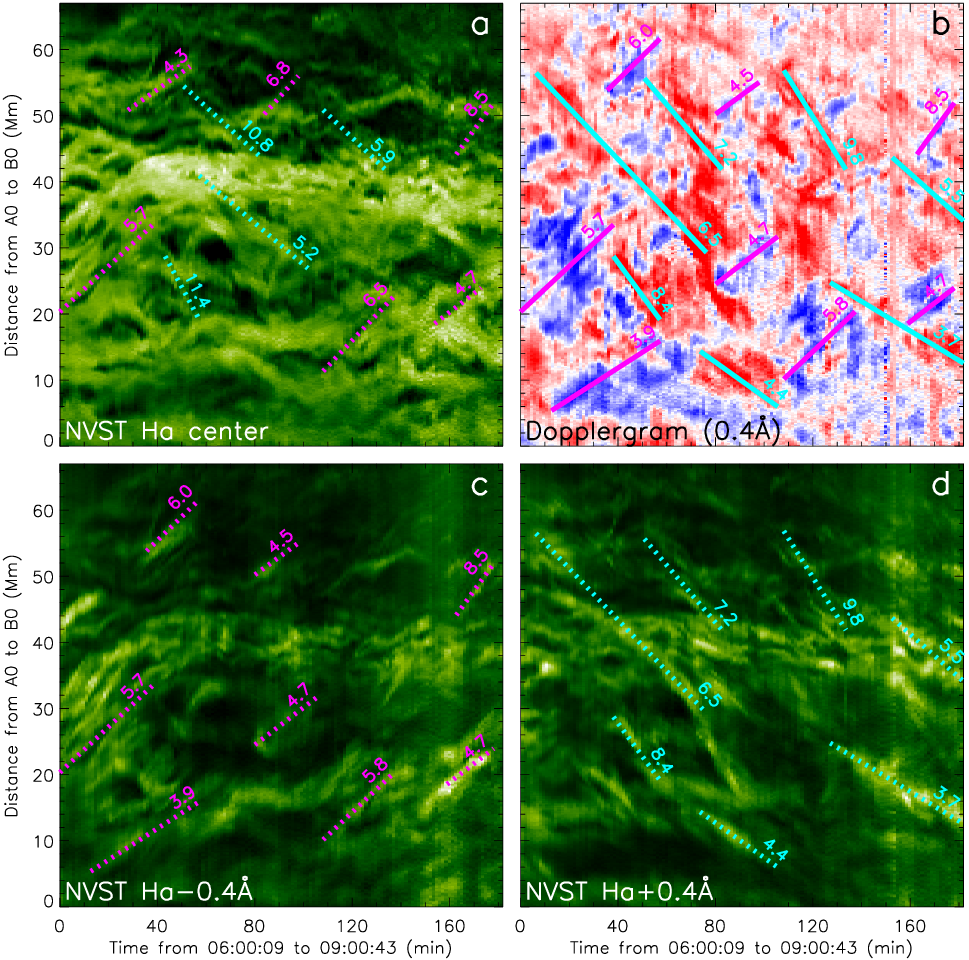}
\caption{{\bf Time-distance diagrams showing the counterstreaming flows along the middle part of the prominence spine.} Panel (a): A time-distance diagram showing the counterstreaming seen in H$\alpha$ line center images. Panel (b): A time-distance diagram showing the counterstreaming seen in Dopplergrams at 0.4 \AA. The red and blue colors denote the redshifted and blueshifted Doppler signals. Panel (c): A time-distance diagram showing the counterstreaming seen in H$\alpha$ off-band -0.4 \AA\ images. Panel (d): A time-distance diagram showing the counterstreaming seen in H$\alpha$ off-band 0.4 \AA\ images. Note that the time-distance diagrams are made along the slit A0-B0 marked in Figure 1e. The pink lines and the cyan lines indicate the opposite directions of flows in the prominence.}
       \label{fig2}
   \end{figure*}
 
   \begin{figure*}
  \centering
   \includegraphics[width=16cm]{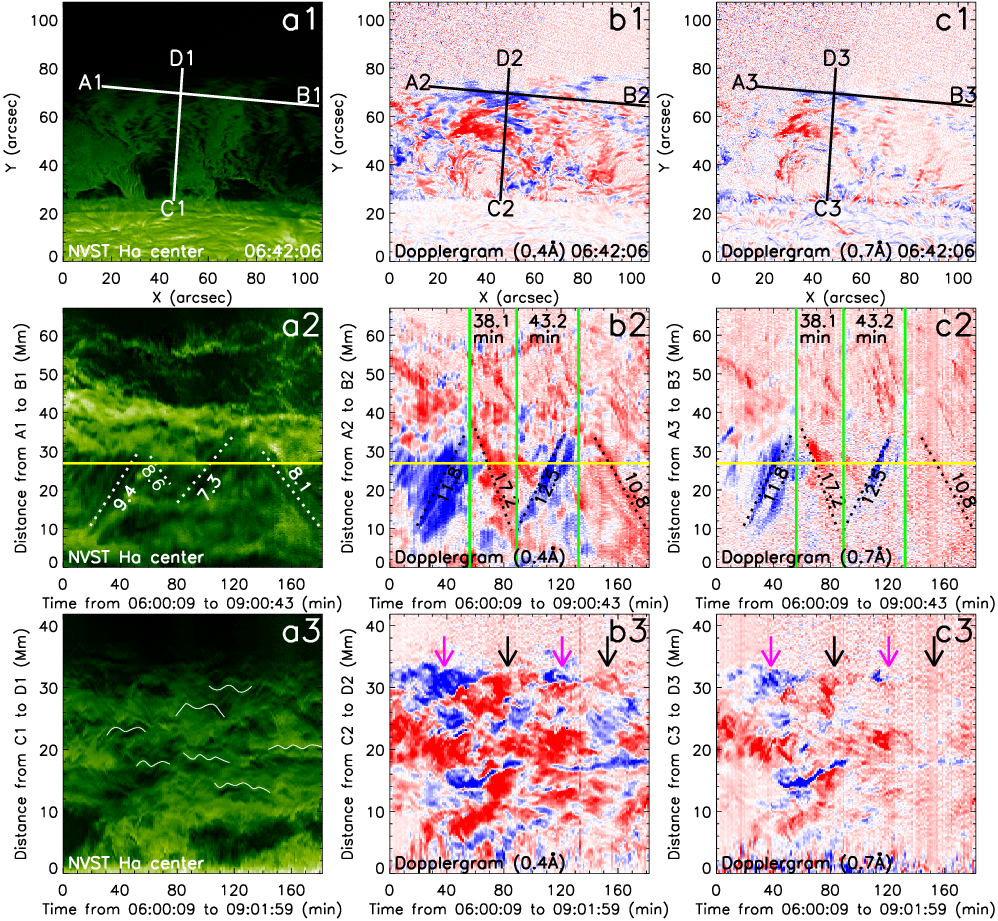}
\caption{{\bf Oscillation of the prominence mass.} Panels (a1-c1): an H$\alpha$ center image, Dopplergrams at 0.4 \AA\ and 0.7 \AA\ at 06:42:06 UT, respectively. The field of view of Figure 2a1 is signed by the red box in Figure 1e. Panels (a2-c2): Time-distance diagrams along slits A1-B1, A2-B2, and A3-B3. The white dotted lines in panel (a2) and the black dotted lines in panels (b2) and (c2) denote the linear fitting along the different features to obtain the velocities of the mass flows along these slits. The vertical green lines separate the red shift from the blue shift in panels (b2) and (c2). The yellow lines are used to create the intensity profiles of panels 3a-3c. Panels (a3-c3): Time-distance diagrams along slits C1-D1, C2-D2, and C3-D3. The curved lines in panel (a3) delineate the small amplitude oscillation of the prominence mass. The pink and black arrows denote the alternant appearance of the blue and the red shifts at the higher part of the prominence spine in panels (b3) and (c3).
}
       \label{fig1}
   \end{figure*}
   
        \begin{figure*}
  \centering
   \includegraphics[width=16cm]{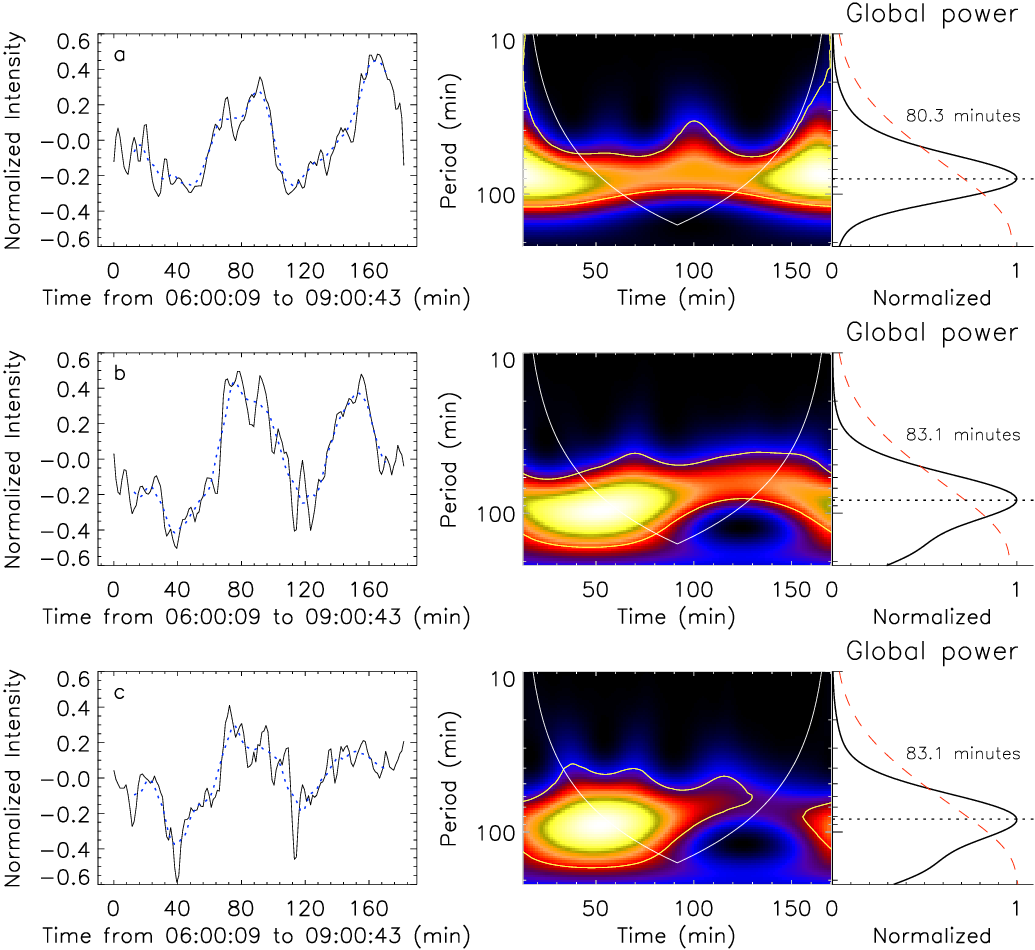}
\caption{{\bf Oscillation period of the prominence mass.} Left panels: The profiles of the normalized intensity (the solid lines) along the yellow lines in Figs. 2a2, 2b2, and 2c2. The dotted lines denote the smoothed intensity profiles with ten points. Right panels: Wavelet analysis for the normalized intensity corresponding to the left panels. The red dash lines indicate the 95\% confidence level.}
       \label{fig3}
   \end{figure*} 

     \begin{figure*}
  \centering
   \includegraphics[width=16cm]{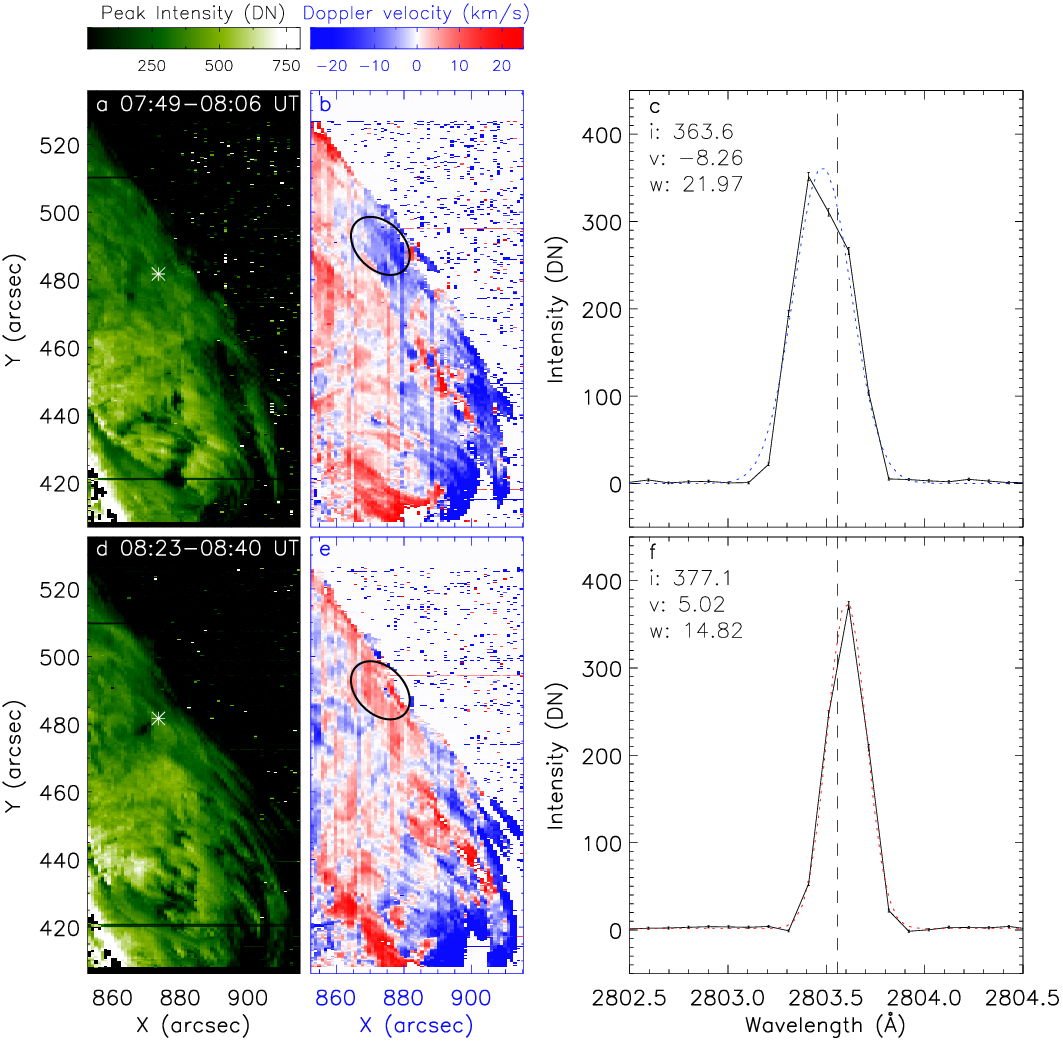}
\caption{{\bf A part of the prominence observed by the IRIS.} Panels (a) and (d): The intensity images acquired at Mg II h 2803.5 \AA. The field of view is marked by the white box in Figures 1b. Panels (b) and (e): The corresponding Doppler velocities derived from this spectral line. It is obviously the Doppler signals in the black circles change from the blueshifted signal to the redshifted one from 07:49 UT to 08:23 UT.  Panels (c) and (f): The profiles of Mg II h (the black solid lines are the observational profiles and the dashed blue and red lines are Gauss fitting profiles.) at the asterisk points in the Figs. 4a and 4d. Note that the the asterisk points are located in the black circle. The word i, v, and w denote the peak intensity, Doppler velocity, and line width, respectively. The field of view of Figs.4a and 4b is outlined by the white box in Figure 1b.}
       \label{fig3}
   \end{figure*} 
   
        \begin{figure*}
  \centering
   \includegraphics[width=16cm]{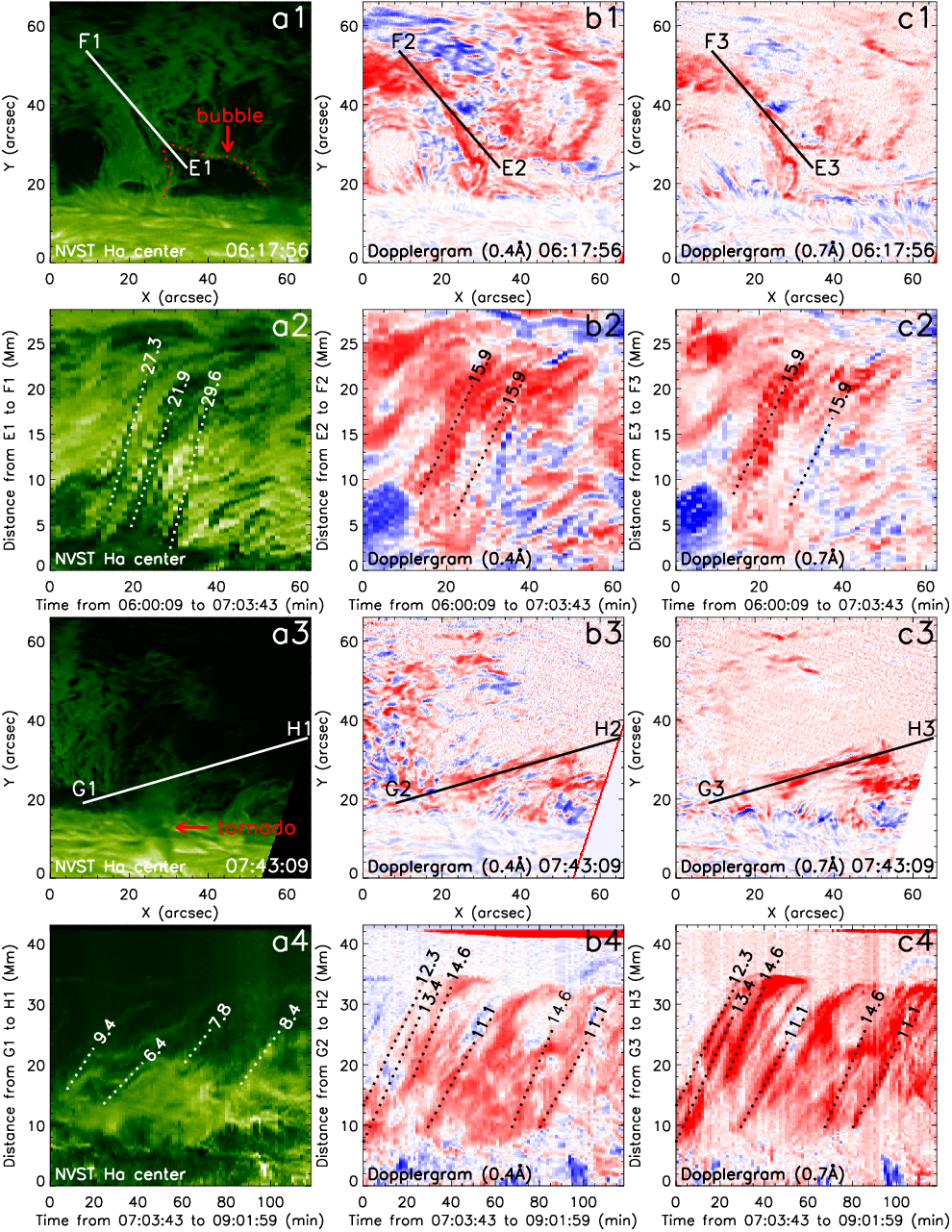}
\caption{{\bf Two types of mass ejection.} Panels (a1-c1): An H$\alpha$ center image, Dopplergrams at 0.4 \AA\ and 0.7 \AA\ at 06:17:56 UT, respectively. The field of view of Figure 3a1 is outlined by the red box in Figure 1f. The curved dotted red lines denote the bubble. Panels (a2-c2): Time-distance diagrams along the slits E1-F1, E2-F2, and E3-F3 in panels a1, b1, and c1. The white and black dotted lines are the linear fitting to obtain the velocities of the mass ejection. Panels (a3-c3): A tornado-like barb in H$\alpha$ center image, Dopplergrams at 0.4 \AA\ and 0.7 \AA\ at 07:43:09 UT, respectively. The field of view of Figure 3a1 is outlined by the cyan box in Figure 1f. Panels (a4-c4): Time-distance diagrams along the slits G1-H1, G2-H2, and G3-H3 in panels a3, b3, and c3. The white and black dotted lines used to derive the projection velocities of the mass ejections.}
       \label{fig4}
   \end{figure*} 
  
    \begin{figure*}
  \centering
   \includegraphics[width=16cm]{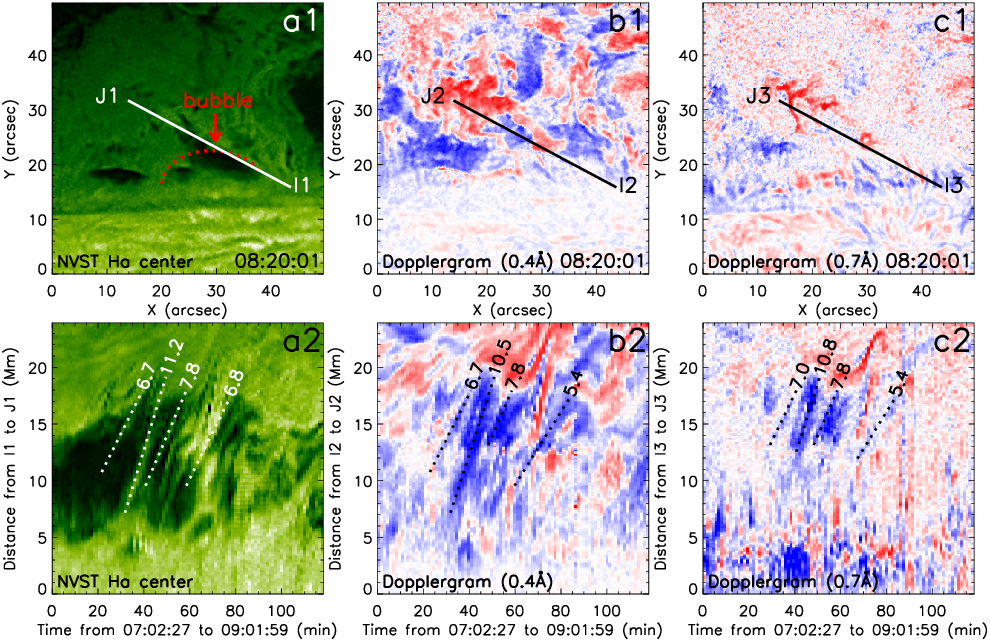}
\caption{{\bf Mass ejection with the upward plume.} Panels (a1-c1): an H$\alpha$ line center image, Dopplergrams at 0.4 \AA\ and 0.7 \AA\ at 08:20:01 UT, respectively. The field of view of Figure 4a1 is signed by the cyan box in Figure 1e. The curved dotted red lines denote the bubble. Panels (a2-c2): Time-distance diagrams along the slits I1-J1, I2-J2, and I3-J3 in panels a1, b1, and c1. The white and black dotted lines are used to derive the projection velocities of the mass ejections.}
       \label{fig2}
   \end{figure*}

\end{CJK*}

\begin{thebibliography}{}
\bibitem[Alexander et al.(2013)]{2013ApJ...775L..32A} Alexander, C.~E., Walsh, R.~W., R{\'e}gnier, S., et al.\ 2013, \apjl, 775, L32. doi:10.1088/2041-8205/775/1/L32
\bibitem[Awasthi et al.(2019)]{2019ApJ...872..109A} Awasthi, A.~K., Liu, R., \& Wang, Y.\ 2019, \apj, 872, 109. doi:10.3847/1538-4357/aafdad
\bibitem[Berger et al.(2008)]{2008ApJ...676L..89B} Berger, T.~E., Shine, R.~A., Slater, G.~L., et al.\ 2008, \apjl, 676, L89. doi:10.1086/587171
\bibitem[Berger et al.(2010)]{2010ApJ...716.1288B} Berger, T.~E., Slater, G., Hurlburt, N., et al.\ 2010, \apj, 716, 1288. doi:10.1088/0004-637X/716/2/1288
\bibitem[Berger et al.(2011)]{2011Natur.472..197B} Berger, T., Testa, P., Hillier, A., et al.\ 2011, \nat, 472, 197. doi:10.1038/nature09925
\bibitem[Bi et al.(2020)]{2020ApJ...891L..40B} Bi, Y., Yang, B., Li, T., et al.\ 2020, \apjl, 891, L40. doi:10.3847/2041-8213/ab79a2
\bibitem[Chen et al.(2008)]{2008A&A...484..487C} Chen, P.~F., Innes, D.~E., \& Solanki, S.~K.\ 2008, \aap, 484, 487. doi:10.1051/0004-6361:200809544
\bibitem[Chen et al.(2014)]{2014ApJ...784...50C} Chen, P.~F., Harra, L.~K., \& Fang, C.\ 2014, \apj, 784, 50. doi:10.1088/0004-637X/784/1/50
\bibitem[Chen et al.(2020)]{2020RAA....20..166C} Chen, P.-F., Xu, A.-A., \& Ding, M.-D.\ 2020, Research in Astronomy and Astrophysics, 20, 166. doi:10.1088/1674-4527/20/10/166
\bibitem[Cai et al.(2024)]{2024ApJ...977..186C} Cai, Y., Xiang, Y., \& Ji, K.\ 2024, \apj, 977, 186. doi:10.3847/1538-4357/ad9006
\bibitem[Deng et al.(2002)]{2002SoPh..209..153D} Deng, Y., Lin, Y., Schmieder, B., et al.\ 2002, \solphys, 209, 153. doi:10.1023/A:1020924406991
\bibitem[Dud{\'\i}k et al.(2012)]{2012ApJ...761....9D} Dud{\'\i}k, J., Aulanier, G., Schmieder, B., et al.\ 2012, \apj, 761, 9. doi:10.1088/0004-637X/761/1/9
\bibitem[De Pontieu et al.(2014)]{2014SoPh..289.2733D} De Pontieu, B., Title, A.~M., Lemen, J.~R., et al.\ 2014, \solphys, 289, 2733. doi:10.1007/s11207-014-0485-y
\bibitem[Diercke et al.(2018)]{2018A&A...611A..64D} Diercke, A., Kuckein, C., Verma, M., et al.\ 2018, \aap, 611, A64. doi:10.1051/0004-6361/201730536
\bibitem[Jing et al.(2006)]{2006SoPh..236...97J} Jing, J., Lee, J., Spirock, T.~J., et al.\ 2006, \solphys, 236, 97. doi:10.1007/s11207-006-0126-1
\bibitem[Ji et al.(2019)]{2019ChSBu..46...1738} Ji, K., Liu, H., Jin, Z., Shang, Z., \& Qiang, Z. 2019, Chinese Science Bulletin, 64, 1738. doi: 10.1360/N972019-00092
\bibitem[Lin et al.(2003)]{2003SoPh..216..109L} Lin, Y., Engvold, O. rn ., \& Wiik, J.~E.\ 2003, \solphys, 216, 109. doi:10.1023/A:1026150809598
\bibitem[Lin et al.(2005)]{2005SoPh..226..239L} Lin, Y., Engvold, O., der Voort, L.~R. van ., et al.\ 2005, \solphys, 226, 239. doi:10.1007/s11207-005-6876-3
\bibitem[Langangen et al.(2008)]{2008ApJ...673.1201L} Langangen, {\O}., Rouppe van der Voort, L., \& Lin, Y.\ 2008, \apj, 673, 1201. doi:10.1086/524057
\bibitem[Lemen et al.(2012)]{2012SoPh..275...17L} Lemen, J.~R., Title, A.~M., Akin, D.~J., et al.\ 2012, \solphys, 275, 17. doi:10.1007/s11207-011-9776-8
\bibitem[Li et al.(2012)]{2012ApJ...752L..22L} Li, X., Morgan, H., Leonard, D., et al.\ 2012, \apjl, 752, L22. doi:10.1088/2041-8205/752/2/L22
\bibitem[Luna et al.(2014)]{2014ApJ...785...79L} Luna, M., Knizhnik, K., Muglach, K., et al.\ 2014, \apj, 785, 79. doi:10.1088/0004-637X/785/1/79
\bibitem[Liu et al.(2014)]{2014RAA....14..705L} Liu, Z., Xu, J., Gu, B.-Z., et al.\ 2014, Research in Astronomy and Astrophysics, 14, 705-718. doi:10.1088/1674-4527/14/6/009
\bibitem[Li et al.(2018)]{2018ApJ...863..192L} Li, D., Shen, Y., Ning, Z., et al.\ 2018a, \apj, 863, 192. doi:10.3847/1538-4357/aad33f
\bibitem[Li et al.(2018)]{2018Ap&SS.363..118L} Li, H., Liu, Y., Tam, K.~V., et al.\ 2018b, \apss, 363, 118. doi:10.1007/s10509-018-3342-x
\bibitem[Ning et al.(2009)]{2009A&A...499..595N} Ning, Z., Cao, W., Okamoto, T.~J., et al.\ 2009a, \aap, 499, 595. doi:10.1051/0004-6361/200810853
\bibitem[Ning et al.(2009)]{2009ApJ...707.1124N} Ning, Z., Cao, W., \& Goode, P.~R.\ 2009b, \apj, 707, 1124. doi:10.1088/0004-637X/707/2/1124
\bibitem[Okamoto et al.(2007)]{2007Sci...318.1577O} Okamoto, T.~J., Tsuneta, S., Berger, T.~E., et al.\ 2007, Science, 318, 1577. doi:10.1126/science.1145447
\bibitem[Okamoto et al.(2015)]{2015ApJ...809...71O} Okamoto, T.~J., Antolin, P., De Pontieu, B., et al.\ 2015, \apj, 809, 71. doi:10.1088/0004-637X/809/1/71
\bibitem[Priest(1989)]{1989ASSL..150....1P} Priest, E.~R.\ 1989, Dynamics and Structure of Quiescent Solar Prominences, 150, 1. doi:10.1007/978-94-009-3077-3\_1
\bibitem[Pesnell et al.(2012)]{2012SoPh..275....3P} Pesnell, W.~D., Thompson, B.~J., \& Chamberlin, P.~C.\ 2012, \solphys, 275, 3. doi:10.1007/s11207-011-9841-3
\bibitem[Parenti(2014)]{2014LRSP...11....1P} Parenti, S.\ 2014, Living Reviews in Solar Physics, 11, 1. doi:10.12942/lrsp-2014-1
\bibitem[Panesar et al.(2013)]{2013A&A...549A.105P} Panesar, N.~K., Innes, D.~E., Tiwari, S.~K., et al.\ 2013, \aap, 549, A105. doi:10.1051/0004-6361/201220503
\bibitem[Panesar et al.(2014)]{2014SoPh..289.2971P} Panesar, N.~K., Innes, D.~E., Schmit, D.~J., et al.\ 2014, \solphys, 289, 2971. doi:10.1007/s11207-014-0504-z
\bibitem[Panesar et al.(2020)]{2020ApJ...897L...2P} Panesar, N.~K., Tiwari, S.~K., Moore, R.~L., et al.\ 2020, \apjl, 897, L2. doi:10.3847/2041-8213/ab9ac1
\bibitem[Panasenco et al.(2014)]{2014SoPh..289..603P} Panasenco, O., Martin, S.~F., \& Velli, M.\ 2014, \solphys, 289, 603. doi:10.1007/s11207-013-0337-1
\bibitem[Su et al.(2012)]{2012ApJ...756L..41S} Su, Y., Wang, T., Veronig, A., et al.\ 2012, \apjl, 756, L41. doi:10.1088/2041-8205/756/2/L41
\bibitem[Schmieder et al.(1991)]{1991A&A...252..353S} Schmieder, B., Raadu, M.~A., \& Wiik, J.~E.\ 1991, \aap, 252, 353
\bibitem[Schmieder et al.(2008)]{2008SoPh..247..321S} Schmieder, B., Bommier, V., Kitai, R., et al.\ 2008, \solphys, 247, 321. doi:10.1007/s11207-007-9100-9
\bibitem[Schmieder et al.(2010)]{2010A&A...514A..68S} Schmieder, B., Chandra, R., Berlicki, A., et al.\ 2010, \aap, 514, A68. doi:10.1051/0004-6361/200913477
\bibitem[Schmieder et al.(2017)]{2017A&A...597A.109S} Schmieder, B., Mein, P., Mein, N., et al.\ 2017, \aap, 597, A109. doi:10.1051/0004-6361/201628771
\bibitem[Shen et al.(2015)]{2015ApJ...814L..17S} Shen, Y., Liu, Y., Liu, Y.~D., et al.\ 2015, \apjl, 814, L17. doi:10.1088/2041-8205/814/1/L17
\bibitem[Song et al.(2024)]{2024ApJ...975..280S} Song, Y., Ning, Z., Li, D., et al.\ 2024, \apj, 975, 280. doi:10.3847/1538-4357/ad813c
\bibitem[Tandberg-Hanssen(1995)]{1995ASSL..199.....T} Tandberg-Hanssen, E.\ 1995, Astrophysics and Space Science Library, vol. 199, Dordrecht: Kluwer Academic Publishers, |c1995. doi:10.1007/978-94-017-3396-0
\bibitem[Vial \& Engvold(2015)]{2015ASSL..415.....V} Vial, J.-C. \& Engvold, O.\ 2015, Solar Prominences, 415. doi:10.1007/978-3-319-10416-4
\bibitem[Wang et al.(2018)]{2018ApJ...852L..18W} Wang, H., Liu, R., Li, Q., et al.\ 2018, \apjl, 852, L18. doi:10.3847/2041-8213/aaa2f4
\bibitem[Wang et al.(2022)]{2022A&A...659A..76W} Wang, J., Yan, X., Xue, Z., et al.\ 2022, \aap, 659, A76. doi:10.1051/0004-6361/202142584
\bibitem[Xiang et al.(2016)]{2016NewA...49....8X} Xiang, Y.-. yuan ., Liu, Z., \& Jin, Z.-. yu .\ 2016, \na, 49, 8. doi:10.1016/j.newast.2016.05.002
\bibitem[Xue et al.(2021)]{2021RAA....21..222X} Xue, J.-C., Vial, J.-C., Su, Y., et al.\ 2021, Research in Astronomy and Astrophysics, 21, 222. doi:10.1088/1674-4527/21/9/222
\bibitem[Yan et al.(2015)]{2015RAA....15.1725Y} Yan, X.-L., Xue, Z.-K., Xiang, Y.-Y., et al.\ 2015, Research in Astronomy and Astrophysics, 15, 1725. doi:10.1088/1674-4527/15/10/009
\bibitem[Yan et al.(2020)]{2020Sci. China Technol. Sci.63...1656} Yan, X., Liu, Z., Zhang, J., et al.\ 2020, SCIENCE CHINA Technological Sciences, 63, 1656
\bibitem[Zirker et al.(1998)]{1998Natur.396..440Z} Zirker, J.~B., Engvold, O., \& Martin, S.~F.\ 1998, \nat, 396, 440. doi:10.1038/24798
\bibitem[Zhang et al.(2012)]{2012A&A...542A..52Z} Zhang, Q.~M., Chen, P.~F., Xia, C., et al.\ 2012, \aap, 542, A52. doi:10.1051/0004-6361/201218786
\bibitem[Zhang et al.(2017)]{2017ApJ...851...47Z} Zhang, Q.~M., Li, D., \& Ning, Z.~J.\ 2017, \apj, 851, 47. doi:10.3847/1538-4357/aa9898
\bibitem[Zhang(2020)]{2020A&A...642A.159Z} Zhang, Q.~M.\ 2020, \aap, 642, A159. doi:10.1051/0004-6361/202038557
\bibitem[Zhang et al.(2024)]{2024MNRAS.533.3255Z} Zhang, Q.~M., Lin, M.~S., Yan, X.~L., et al.\ 2024, \mnras, 533, 3255. doi:10.1093/mnras/stae1936
\bibitem[Zhou et al.(2020)]{2020NatAs...4..994Z} Zhou, Y.~H., Chen, P.~F., Hong, J., et al.\ 2020, Nature Astronomy, 4, 994. doi:10.1038/s41550-020-1094-3
\bibitem[Zhou et al.(2021)]{2021A&A...647A.112Z} Zhou, C., Xia, C., \& Shen, Y.\ 2021, \aap, 647, A112. doi:10.1051/0004-6361/202039558




\end{thebibliography}
\end{document}